# First-Principles Demonstration of Non-adiabatic Thouless Pumping of Electrons in A Molecular System

Ruiyi Zhou[1], Dillon C. Yost[1,2], and Yosuke Kanai[1]
1. Department of Chemistry, University of North Carolina at Chapel Hill, NC.
2. Department of Materials Science and Engineering , Massachusetts Institute of Technology, MA.

**Abstract**
We demonstrate non-adiabatic Thouless pumping of electrons in *trans*-polyacetylene in the framework of Floquet engineering using first-principles theory. We identify the regimes in which the quantized pump is operative with respect to the driving electric field for a time-dependent Hamiltonian. By employing the time-dependent maximally-localized Wannier functions in real-time time-dependent density functional theory simulation, we connect the winding number, a topological invariant, to a molecular-level understanding of the quantized pumping. While the pumping dynamics constitutes the opposing movement of the Wannier functions that represent both double and single bonds, the resulting current is unidirectional due to the greater number of double bond electrons. Using a gauge-invariant formulation called dynamical transition orbitals, an alternative viewpoint on the non-equilibrium dynamics is obtained in terms of the particle-hole excitation. A *single* time-dependent transition orbital is found to be largely responsible for the observed quantized pumping. In this representation, the pumping dynamics manifest itself in the dynamics of this single orbital as it undergoes changes from its $\pi$ bonding orbital character at equilibrium to acquiring resonance and anti-bonding character in the driving cycle. The work demonstrates the Floquet engineering of the non-adiabatic topological state in an extended molecular system, paving the way for experimental realization of the new quantum material phase.

-----------------

The quantized topological pumping phenomenon was first proposed by Thouless[1] in 1983, now widely referred to as Thouless pumping. Thouless's seminal work on the quantized particle transport in a slow varying potential showed that the quantization of the number of pumped particles derives from topology of the underlying quantum-mechanical Hamiltonian, given by a Chern number. In recent years, Thouless pumping has been demonstrated experimentally in various systems[2-3] including an ultracold Fermi gas[4] and ultracold atoms in optical lattice[5]. Most theoretical studies have employed model Hamiltonians[6-11] such as the Rice-Mele model[12] and the description of topological pumping had assumed a complete adiabaticity of the Hamiltonian evolution. More recent work has begun to study the non-adiabatic effect in Thouless pumping[13-15]. In addition to studying the non-adiabatic effect on the otherwise adiabatic Thouless pump, non-adiabatic variance of Thouless pump has been proposed through setting up a periodically-driven system. The idea of this so-called topological Floquet engineering is to use a time-periodic field to induce the topological properties in a driven system that is otherwise a trivial insulator[16-17]. In a Floquet system, time-dependent Hamiltonian satisfies $\widehat{H}(t + T) = \widehat{H}(t)$ and time-independent effective Hamiltonian can be defined from the evolution operator over one periodic $T$ such that $\widehat{H}_{eff}(\boldsymbol{k}) \equiv iT^{-1}\ln\left\{\widehat{\mathcal{T}}exp\left[-i\int_0^T dt\,\widehat{H}(\boldsymbol{k},t)\right]\right\}$. The term Floquet band engineering is often used to describe a theoretical approach that changes topological properties of a system through the energy spectrum of the effective Hamiltonian. As discussed in the recent review by Rudner and Lindner[17], for example, the topological Floquet band engineering aims to induce the non-trivial insulator phase by the use of a periodic driving field, whereas the system is in the trivial insulator phase without such a driving field. Non-equilibrium topological phenomena thus can be induced by applying time-periodic fields to Hamiltonian. Nakagawa, et al. studied different types of Floquet topological states for the periodically-driven system in the non-adiabatic regime, and a non-adiabatic Thouless pump falls under gapless Floquet topological state of Class A in one dimension[18]. The winding number is equal to the integrated particle current over the periodic time $T$. The topological invariant is given in terms of the energy spectrum of the effective Floquet Hamiltonian, $\varepsilon_i$ (quasienergy[19]), or equivalently in terms of the geometric phase

$$W = \int_{BZ} dk \sum_i^{Occ.} \frac{\partial \varepsilon_i(k)}{\partial k} = -\frac{1}{2\pi} \int_{BZ} dk \sum_i^{Occ.} \frac{\partial \gamma_i^{NA}}{\partial k} \tag{1}$$

where the non-adiabatic Aharonov-Anandan geometric phase[20] of the Floquet states $\Phi_i(k,t)$ is

$$\gamma_i^{NA}(k) \equiv \int_0^T dt \, \langle \Phi_i(k,t) | i\partial_t | \Phi_i(k,t) \rangle . \tag{2}$$

Building on the work by Kitagawa, et al.[21], Nakagawa, et al. showed that the winding number is also expressed as

$$W = \int_{BZ} dk \sum_i^{Occ.} \langle \Phi_i(k,t=0) | \hat{U}^\dagger(k) i\partial_k \hat{U}(k) | \Phi_i(k,t=0) \rangle \tag{3}$$

where $\hat{U}(k)$ is the Floquet operator (i.e. $\hat{U}(k) \equiv \hat{T} exp\left[-i \int_0^T dt \, \hat{H}(k,t)\right]$)[18]. This can be equivalently expressed in terms of the more familiar time-dependent Berry phase,

$$W = \frac{1}{2\pi} \sum_i^{Occ.} \left[ \int_{BZ} dk \, \langle u_i(k,t=T) | i\partial_k | u_i(k,t=T) \rangle - \int_{BZ} dk \, \langle u_i(k,t=0) | i\partial_k | u_i(k,t=0) \rangle \right] \tag{4}$$

where $u_i(k,t)$ is the periodic part of the single-particle Bloch wavefunctions for an extended periodic system. Note that the Floquet state $\Phi_i(k,t)$ and $u_i(k,t)$ are related by $|u_i(k,t)\rangle = e^{-i\varepsilon_i(k)t}|\Phi_i(k,t)\rangle$. For Floquet topological states of a periodically-driven system, the winding number is a non-zero integer, and the winding number can be also expressed, analogously to the static Chern insulator, as

$$W = C \equiv \frac{1}{2\pi} \int_0^T dt \int_{BZ} dk \sum_i^{Occ.} F_i(k,t) \tag{5}$$

where $C$ is the first Chern number of the Floquet states, and the generalized Berry curvature is given by $F_i(k,t) = i[\langle \partial_t u_i(k,t) | \partial_k u_i(k,t) \rangle - \langle \partial_k u_i(k,t) | \partial_t u_i(k,t) \rangle]$[22]. The Berry phase formulation is directly connected to the time-dependent Wannier functions, and the winding number can be interpreted as the number of the geometric centers of the Wannier functions (i.e. Wannier centers) pumped over one driving cycle in the Thouless pumping[18, 23]. Note that the Wannier functions can be defined equivalently in terms of either $u_i(k,t)$ or Floquet states $\Phi_i(k,t)$ because they are related by a phase factor. Wannier centers have already been used in literature for characterizing topological insulators in the adiabatic description[8, 23]. In the recent work by Yost, et al.[24], we introduced the non-adiabatic dynamics of maximally-localized Wannier functions[25-26] in real-time time-dependent density functional theory (RT-TDDFT), and its application to studying Thouless topological charge pumping was briefly discussed. The winding number can be conveniently expressed in terms of the time-dependent maximally localized Wannier functions (MLWFs), $w_i(\mathbf{r}, t)$, as

$$W = \sum_i^{Occ.} \langle w_i(t=T)|\hat{\mathbf{r}}|w_i(t=T)\rangle - \langle w_i(t=0)|\hat{\mathbf{r}}|w_i(t=0)\rangle, \quad (6)$$

due to the Blount identity $\langle w_i(t)|\hat{\mathbf{r}}|w_i(t)\rangle = \frac{1}{2\pi}\int_{BZ} d\mathbf{k}\, \langle u_i(\mathbf{k},t)|i\partial_\mathbf{k}|u_i(\mathbf{k},t)\rangle$ in the first-principles electronic structure theory, and the position operator here is defined according to the formula given by Resta for the extended periodic systems[27]. In this work, we study whether non-adiabatic Thouless pumping, a Floquet topological state, can be observed for electrons in a real molecular system. We demonstrate the non-adiabatic Thouless pumping of electrons in *trans*-polyacetylene polymer, using time-dependent first-principles electronic structure theory. *Trans*-polyacetylene represents a chemical exemplification of the classic Su-Schrieffer-Heeger (SSH) model which is widely used to study the transition between topological insulator and normal/trivial insulator phases by artificially changing the empirical hopping parameters between the single and double C-C bonds in this Peierls distorted system[28]. The two distinct carbon atom sites from the Peierls instability give rise to the chiral (i.e. sublattice) symmetry in the Hamiltonian.

We apply a time-dependent electric field as the driving field with a specific period (i.e. frequency) along the polymer direction. The electron current is directly obtained from the flux of the geometric centers of the time-dependent MLWFs. Computational details are given Computational Method section. In the context of Floquet theory, the initial state must return to the same Hilbert subspace of the Floquet operator after a driving period. This is not the adiabatic evolution condition, and particle-hole excitations are still allowed in this formulation as discussed by Nakagawa, et al.[18]. This aspect is also relevant for discussing the applicability of the Floquet theory to time-dependent density functional theory (TDDFT)[29-32]. The Kohn-Sham Hamiltonian depends on the time-dependent electron density even when the adiabatic approximation is adapted for the exchange-correlation potential and the dependence on the initial state is consequently neglected[33-34]. Therefore, the Hamiltonian is time-periodic (i.e. Floquet theory is applicable) only if the original electron density is recovered after each driving cycle is completed. We can quantify the extent to which the Floquet theorem is satisfied by calculating the determinant of the overlap matrix, S, between the initial time-dependent Kohn-Sham (TD-KS) orbitals and those after one driving cycle has passed as shown in Figure 1. Dark shaded areas with values close to one satisfy this condition and the Floquet theory is applicable, and the integrated current over one driving cycle is equal to the winding number.

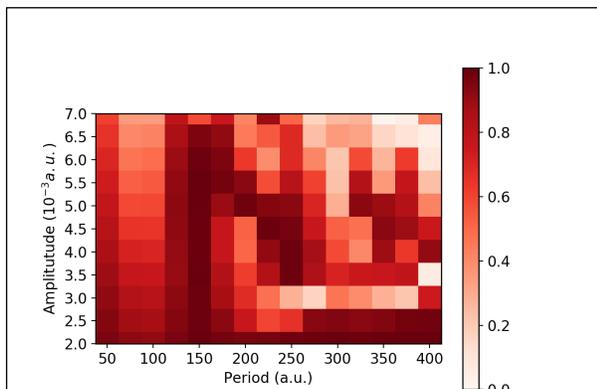

Figure 1: Determinant of the overlap matrix, S, between the initial and final TD-KS states in a single driving cycle as a function of the driving field period and amplitude, sampled at uniform intervals of 25 a.u. and $0.5\times10^{-3}$ a.u., respectively.

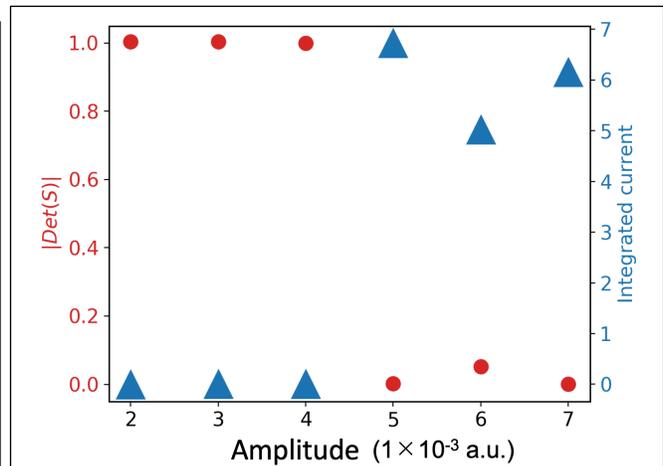

Figure 2: Determinant of the overlap matrix between the initial and final TD-KS states after a single driving cycle (Left) and the integrated current per one C-C monomer unit (Right) for the driving field period of T=1000 a.u. as a function of the driving field amplitude.

In the low frequency regime (e.g. $T=1000$ a.u.) toward the adiabatic evolution limit (Fig. 2), particle-hole excitations are not possible, given the much larger energy gap of 2.02 eV, which corresponds to the electric field with $T=84.6$ a.u.. Below a certain electric field amplitude ($\sim 4\times 10^{-3}$ a.u.), the overlap matrix determinant is essentially one, indicating the original electron density is recovered after one driving cycle as seen in Figure 2. However, this is solely because the electron current is absent. At the same time, when the field amplitude exceeds the threshold, Zener tunneling is possible in the extended systems; it causes the electrical breakdown, resulting in a very large current (Figure 2). The electron density, however, does not return to its initial state after one driving cycle and the overlap matrix determinant is therefore not close to one (see Figure 1). The Zener tunneling is observed above the threshold field amplitude but not the topological pump.

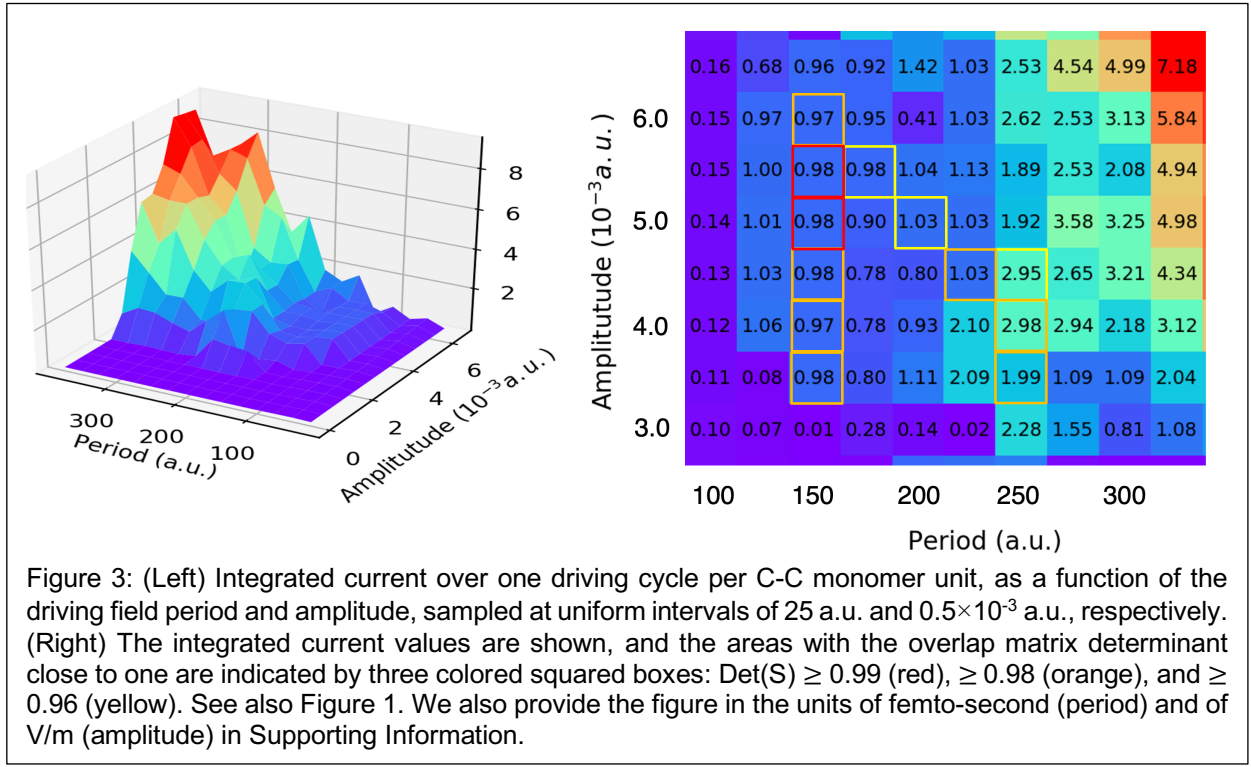

Figure 3: (Left) Integrated current over one driving cycle per C-C monomer unit, as a function of the driving field period and amplitude, sampled at uniform intervals of 25 a.u. and $0.5\times 10^{-3}$ a.u., respectively. (Right) The integrated current values are shown, and the areas with the overlap matrix determinant close to one are indicated by three colored squared boxes: Det(S) ≥ 0.99 (red), ≥ 0.98 (orange), and ≥ 0.96 (yellow). See also Figure 1. We also provide the figure in the units of femto-second (period) and of V/m (amplitude) in Supporting Information.

For various combinations of the period (50~400 a.u.) and amplitude (2~7 $\times 10^{-3}$ a.u.) of the driving field, Figure 3 (Left) shows the integrated electron current per C-C monomer unit over one driving cycle. In order for the integrated current over one driving cycle to be identified as the winding number, the Floquet condition must be satisfied such that the overlap matrix determinant is unity. Figures 1 and 3 show that the Floquet topological state is obtained for some combinations of the driving field period and amplitude. First, the overlap matrix determinant must be one so that the time-dependent electron density returns to its original state after one driving period. This is a necessary condition for the winding number formulation as the Floquet theorem is applied to TDDFT because the Hamiltonian depends on the electron density. At the same time, the winding number itself must be a non-zero value for the Floquet topological state. As indicated by the enclosed squares in Figure 3 (Right), certain combinations of the period and amplitude yield the required features of the Floquet topological state. Since the overlap matrix determinant is never exactly one in numerical simulations, we indicate the areas that show the overlap matrix determinant value of ≥ 0.96 with a non-negligible electron current. For the Floquet topological state indicated by these enclosed squares, the integrated current per driving cycle is given by the winding number and an integer value is expected. While the numerical simulation does not give exactly an integer number here, the computed values show a discrete quantization expected for the integrated current, effectively yielding the winding number of one, two, or three in the areas where the Floquet condition is satisfied.

Having a Floquet topological state identified for this real molecular system with certain driving periods and amplitudes, we now study the nature of the electron current in the Floquet topological state. While the winding number is the physical observable here, orbital analysis of the simulation enables us to gain chemical insights into the dynamics of the non-adiabatic topological pump[35]. As discussed by Nakagawa, et al.[18], the winding number can be interpreted as the number of the Wannier centers pumped over one driving cycle in the Thouless pumping. As can be seen in Figure 4, individual monomer units show two and one MLWFs for the double and single C-C bonds, respectively. Each MLWF represents two electrons of the opposite spins. In the equilibrium ground state, the Wannier function spreads for the double and single C-C bonds are 2.32 and 1.78 a.u.$^2$, respectively. The time-dependent MLWFs remain highly localized; the mean double-bond Wannier spreads are 2.3~3.4 a.u.$^2$, and the single-bond Wannier spreads are 1.7~1.9 a.u.$^2$, depending on the field amplitude and period (see Supporting Information). Let us now discuss the pumping behavior for the winding number of one, $W=1$. As a representative case, we analyze the particular condition with the field period=150 a.u. and amplitude=4 ×10$^{-3}$ a.u., which gives the integrated current of 0.97e numerically (per driving cycle per C-C monomer unit). The time-dependent MWLFs remain highly localized during the non-equilibrium dynamics; the average spread values of the double- and single- bond MLWFs do not increase above 3.0 and 1.9 a.u.$^2$, respectively. Figure 4(a) shows time-dependent changes of the Wannier centers over one period. The MLWFs that correspond to C-H bonds remain essentially unchanged. All the Wannier centers remain oscillating back and forth when the electron current is absent (see Supporting Information). Figure 4 shows that the Wannier center movement for the C-C double and single bonds. While both of the two double-bond Wannier centers move in one direction, the single-bond Wannier centers move in the opposite direction, resulting in the overall directional transport with the winding number of one. At the same time, the MLWF dynamics do not yield a conceptually simple description of a single C-C bond being pumped in one direction as in the SSH model. There are instances at which all MLWFs localize in the vicinity of a carbon atom as seen at $t = 3/8T$ in Figure 4(a).

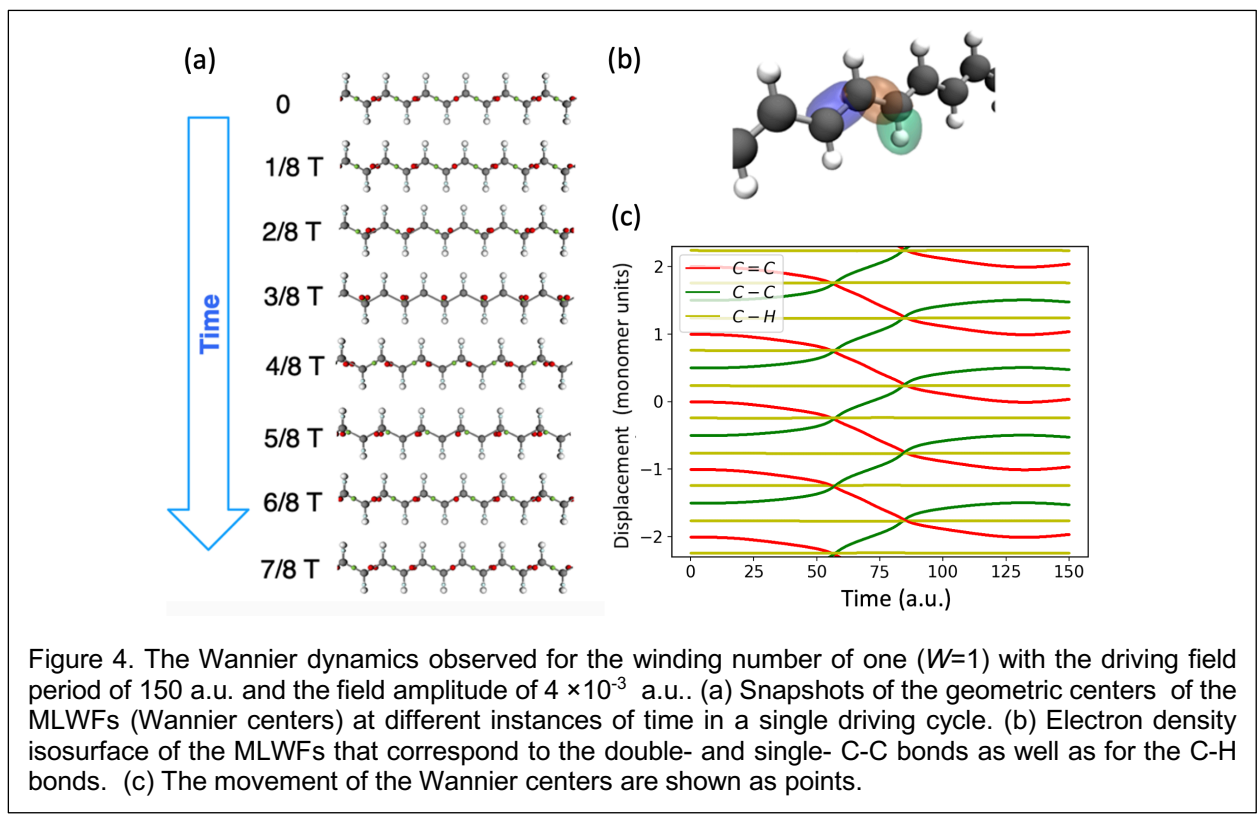

Figure 4. The Wannier dynamics observed for the winding number of one ($W=1$) with the driving field period of 150 a.u. and the field amplitude of 4 ×10$^{-3}$ a.u.. (a) Snapshots of the geometric centers of the MLWFs (Wannier centers) at different instances of time in a single driving cycle. (b) Electron density isosurface of the MLWFs that correspond to the double- and single- C-C bonds as well as for the C-H bonds. (c) The movement of the Wannier centers are shown as points.

The driving field period=250 a.u. with the amplitude=$3.5 \times 10^{-3}$ a.u. yields the winding number of two ($W$=2, 1.99 numerically), and the winding number of three ($W$=3, 2.98 numerically) is obtained with the driving field period=250 a.u. and amplitude=$4 \times 10^{-3}$ a.u. The corresponding Wannier center movement shows that they are pumped over a multiple number of C-C monomer units, given by the winding number (see Fig 5 (a)). The MLWF dynamics still uphold the molecular description such that only single and double bonds are formed alternatively on the C-C units while the winding number itself does not mathematically impose such a physical condition (see Supporting Information). As can be seen in Figure 5 (a), the Wannier center movement shows a behavior that is much more complex than the case for $W$=1 (Fig. 4). The movement of the Wannier centers increases rapidly in the middle of the driving cycle. Particularly for the $W$=3 case, the Wannier centers exhibit a small but abrupt jump around the middle of the driving cycle as evidenced in the time derivative of the Wannier center positions (see Fig. 5 (b)). At the same time, the spreads of the MLWFs remain rather well localized (see Supporting Information). A similar behavior was seen in our earlier work[24] in which a discontinuous movement of the Wannier centers was also observed but in a different context (i.e. optically-gated transistor setup with time-independent homogeneous electric field and optical-excitation field). To summarize our observation here, the winding numbers larger than one, $W$>1, do not imply a simple repetition of the $W$=1 pumping over the C-C monomer units by an integer number times, given by the winding number (e.g. two and three here). This point is further evidenced in the following by studying the pumping dynamics in terms of the particle-hole excitation.

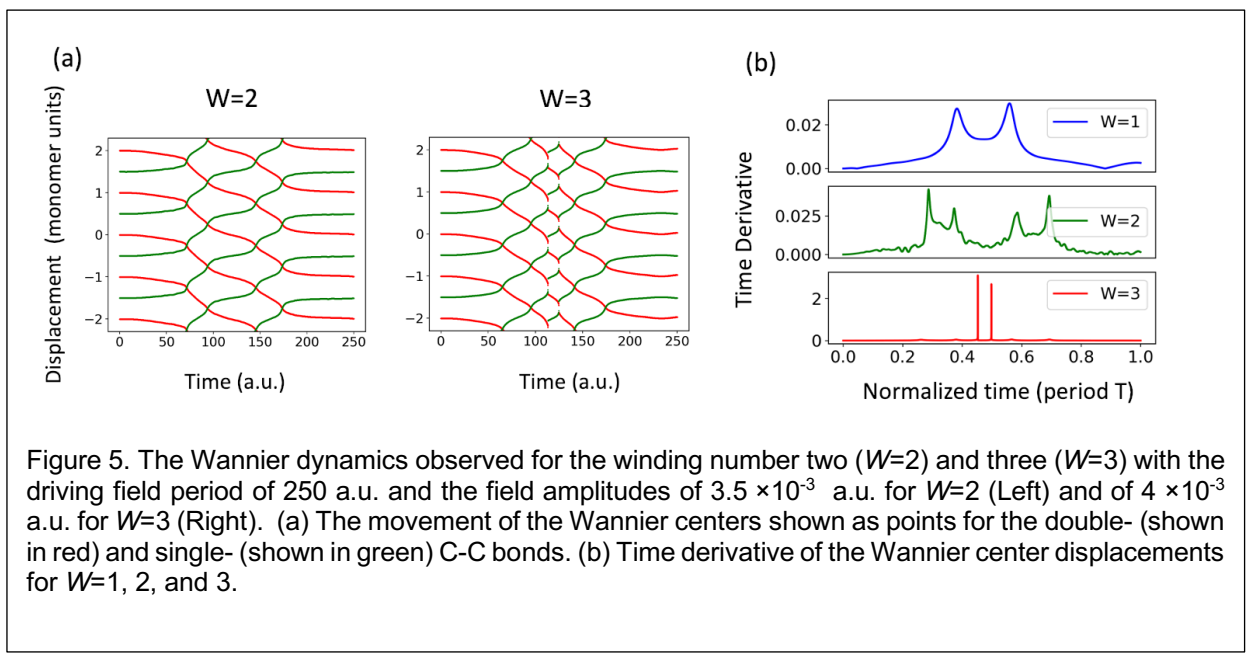

Figure 5. The Wannier dynamics observed for the winding number two ($W$=2) and three ($W$=3) with the driving field period of 250 a.u. and the field amplitudes of $3.5 \times 10^{-3}$ a.u. for $W$=2 (Left) and of $4 \times 10^{-3}$ a.u. for $W$=3 (Right). (a) The movement of the Wannier centers shown as points for the double- (shown in red) and single- (shown in green) C-C bonds. (b) Time derivative of the Wannier center displacements for $W$=1, 2, and 3.

While the winding number formulation through the Wannier function is intuitive due to having the localized description of electrons, an alternative viewpoint of this non-adiabatic Thouless pump can be developed by studying the non-equilibrium dynamics in terms of the particle-hole excitation dynamics. In the recently introduced dynamical transition orbitals (DTO) approach, the RT-TDDFT simulation is framed in a new set of the gauge-invariant time-dependent orbitals[36]. Within this DTO gauge, the particle-hole excitation dynamics is obtained by formulating the individual orbitals a linear combination of the hole and particle orbitals,

$$|\psi_i^{DTO}(t)\rangle = a_i(t)|\psi_i^h(t)\rangle + b_i(t)|\psi_i^p(t)\rangle \quad i=1...N_{Occ.} \tag{7}$$

where the $|\psi_i^h(t)\rangle$ and $|\psi_i^p(t)\rangle$ are the hole and particle orbitals, respectively. The real-valued expansion coefficients satisfy

$$a_i(t) \geq 0, \quad b_i(t) \geq 0, \quad a_i(t)^2 + b_i(t)^2 = 1. \tag{8}$$

The particle and hole orbitals are essentially time-invariant here (up to the phase) while the hole and particle coefficients vary in time. Figure 6 shows the particle population, $b_i(t)^2$, in Eqs. 7/8 for three DTOs with the most dominant changes. Interestingly, a *single* DTO is predominantly responsible for the non-equilibrium dynamics for all the cases. By projecting this particular DTO onto reference Kohn-Sham eigenstates, one finds that the constituting hole and particle orbitals largely derive from degenerate HOMOs and degenerate LUMOs, respectively. Therefore, these particular orbital dynamics, into which the pumping dynamics manifests itself, represents $\pi$ bonds at equilibrium initially at $t=0$. In contrast to the dynamics of MLWFs, a conceptually simple understanding of the Thouless pump can be obtained by studying solely the dynamical change of this particular transition orbital. For the $W=1$ case as a representative example, Figure 7 shows that this orbital evolves from being localized on the typical C-C double bonds as the $\pi$ bonding state into a resonance state before localizing again but on the other alternating set of C-C bonds at $t=0.5T$, exhibiting a significant $\pi\,anti$-bonding character. Then, it continues to evolve into another resonance state before it returns again as the $\pi$-bonding orbital on the C-C double bonds. Pendas, et al. have previously discussed how topological invariant properties can be rationalized in terms of such a chemically intuitive description [37], and a simple schematics can depict the $W=1$ dynamics as shown in Figure 7. Similar DTO dynamics can be observed for the $W=2$ and 3 cases but with more rapid changes in the midst of the driving cycle. This is in accordance with the Wannier dynamics that show more rapid changes in the middle of the driving cycle (Figure 5). As seen in the Wannier center dynamics (Figure 5), the $W=2, 3$ cases do not represent a simple repetition of the $W=1$ transition orbital changes. Indeed, all the three cases show varying dynamics for the transition orbital in terms of the particle occupation change in a single driving cycle (Figure 6). For $W=2$, similarly to $W=1$ case, the transition orbital gains resonance character before $t=0.3T$, anti-bonding character at $t=0.3T$, and then another resonance character at $t=0.4T$ (Figure 7). The transition orbital then repeats the visually similar changes in the second half of the driving cycle. The $W=3$ case is qualitatively dissimilar to the $W=1$ or $W=2$ cases. The transition orbital gains the resonance character at $t=0.2T$, and anti-boding character at $t=0.3T$. However, change to another resonance state character is not observed until much later, $t=0.7T$. The transition orbital undergoes visibly small changes for $t=0.3\sim0.7T$ in the driving cycle, and the fact that the winding number is three is not obvious or intuitive from the dynamical changes of this transition orbital. Although the winding number can be interpreted as the number of the Wannier centers pumped over one driving cycle[18], the changes in the dynamical transition orbital cannot be interpreted straightforwardly in relation to the winding number. At the same time, in this dynamical transition orbital gauge, a single orbital effectively captures the pump dynamics, and its changes are more intuitive from the viewpoint of understanding the dynamics in terms of electron transition among the familiar chemical bonding orbitals.

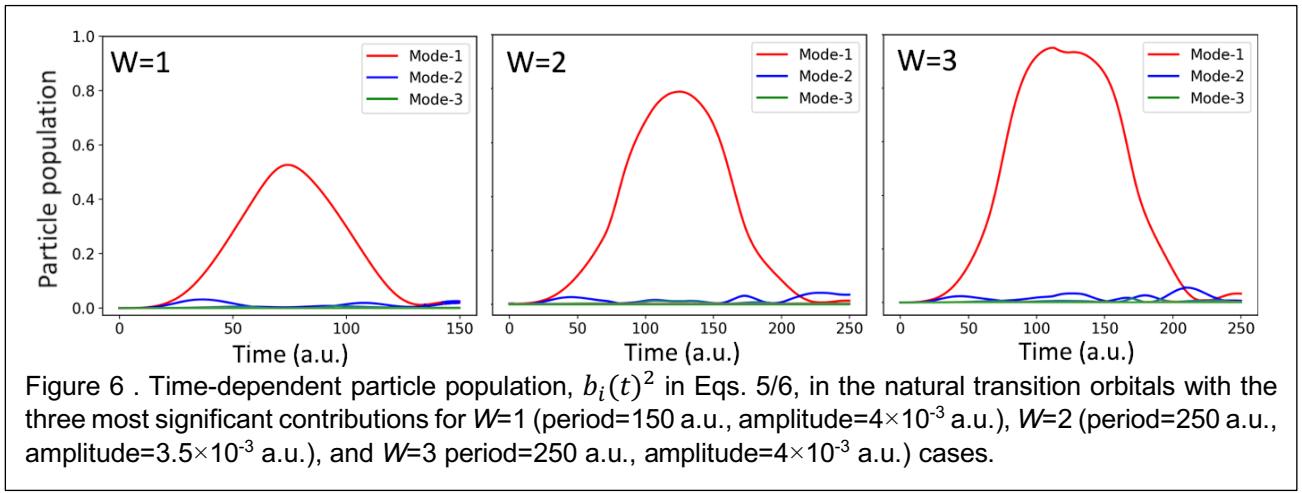

Figure 6. Time-dependent particle population, $b_i(t)^2$ in Eqs. 5/6, in the natural transition orbitals with the three most significant contributions for $W=1$ (period=150 a.u., amplitude=4×10⁻³ a.u.), $W=2$ (period=250 a.u., amplitude=3.5×10⁻³ a.u.), and $W=3$ period=250 a.u., amplitude=4×10⁻³ a.u.) cases.

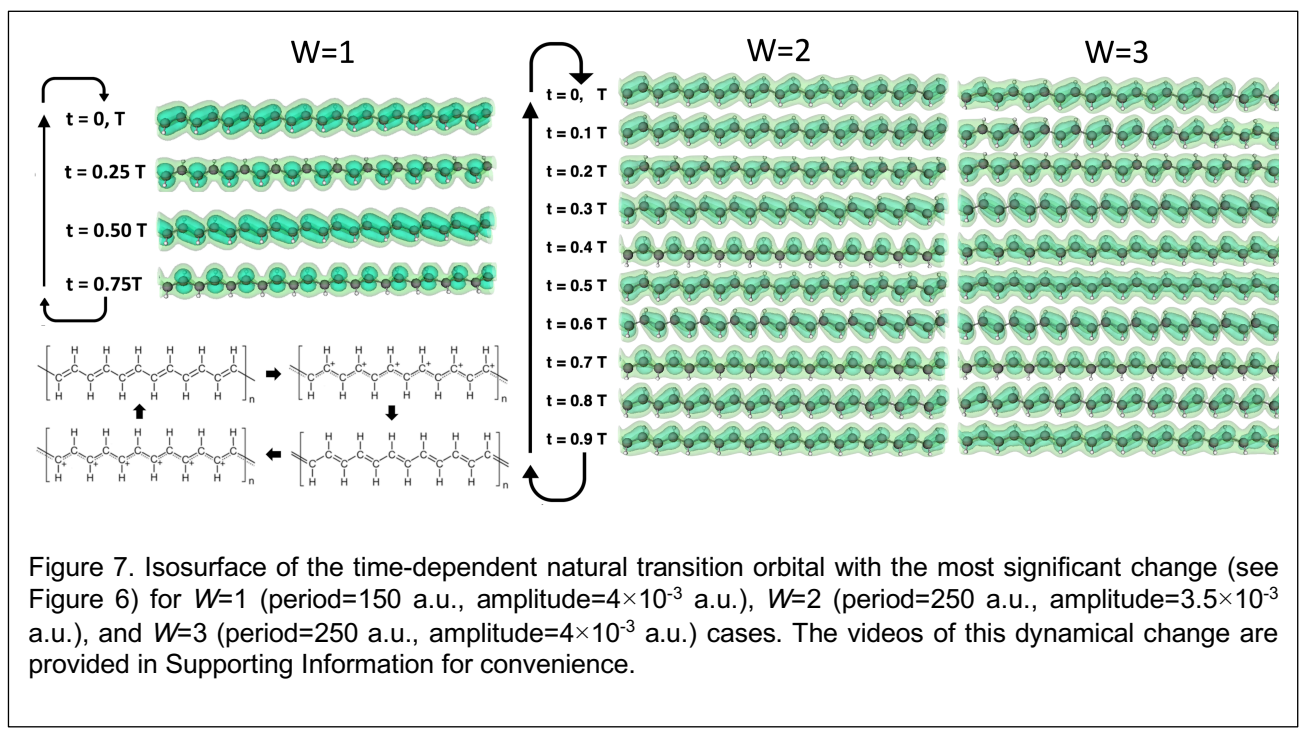

Figure 7. Isosurface of the time-dependent natural transition orbital with the most significant change (see Figure 6) for $W=1$ (period=150 a.u., amplitude=$4\times10^{-3}$ a.u.), $W=2$ (period=250 a.u., amplitude=$3.5\times10^{-3}$ a.u.), and $W=3$ (period=250 a.u., amplitude=$4\times10^{-3}$ a.u.) cases. The videos of this dynamical change are provided in Supporting Information for convenience.

Using first-principles, time-dependent, electronic structure theory, we demonstrated non-adiabatic Thouless pumping of electrons in *trans*-polyacetylene in the framework of Floquet engineering. Using an external electric field as the driving field for time-dependent Hamiltonian, we identified the regimes in which the quantized pump is operative as a function of the driving field period (frequency) and amplitude. By employing the time-dependent maximally-localized Wannier functions in real-time time-dependent density functional theory simulations, we connect the winding number formulation by Nakagawa, et al.[18] to a molecular-level understanding of this Thouless pump. These pumping dynamics are described by the movement of the Wannier functions that represent both C-C double bonds and single bonds but not C-H bonds. While the Wannier centers movein opposite directions, having more electrons in the double bonds than in the single bonds results in a net unidirectional current overall. The direction of the current is governed by the initial polarization direction of the applied time-dependent electric field. Although the topological invariant, the winding number, specifies the number of C-C monomer units the electrons are pumped over in a single driving cycle, the Wannier dynamics shows that the rate at which the electrons are pumped is not uniform during the driving cycle for the cases with larger winding numbers of two and three. Using a gauge-invariant formulation of the occupied time-dependent Kohn-Sham orbitals called dynamical transition orbitals[36], a *single* time-dependent orbital is found to be responsible for the observed Thouless pumping. The pumping dynamics manifest in the dynamic changes of this single transition orbital that characterizes the $\pi$ bonding at equilibrium, and it undergoes changes to acquire resonance and anti-bonding character in the driving cycle. Having the dynamics formulated in terms of the particle-hole transition, it was further shown that the larger winding numbers of two and three do not indicate a simple repetition of the $W$=1 pumping over C-C monomer units by two and three times as the electrons are pumped over two and three C-C monomer units, respectively. Topological insulators are generally studied in terms of electronic structure in a static field of the fixed-position nuclei; how lattice dynamics of nuclei influence the exotic transport behavior at room temperature is of great importance for experimental realization and for practical application of quantum materials. This aspect is particularly relevant for the present case because the topological properties are not guaranteed to be robust in the non-adiabatic regime, including Floquet topological states[17]. Investigation into the lattice dynamics and thermal effects will be pursued in a future work.

## Computational Details

Real-time time-dependent functional theory (RT-TDDFT)[29, 38] has become an increasingly popular method for studying out-of-equilibrium electronic structures in the past few decades[39-40], including those of topological materials[24, 41-42]. RT-TDDFT simulation was performed using the Qb@ll branch[43] of Qbox code[44] within plane wave pseudopotentials (PW-PP) formalism[45]. We employed a 55-atom supercell (11 C-C monomer unit cells aligned along the x axis) with the periodic boundary condition (51.32 Bohr ×15.0 Bohr × 15.0 Bohr) and the Γ-point approximation in the Brillouin zone integration. Molecular geometry (bond lengths, bond angles, and the lattice constant) of the *trans*-polyacetylene was taken from that of experiments[46] and included in Supporting Information. All atoms were represented by Hamann-Schluter-Chiang-Vanderbilt (HSCV) norm-conserving pseudopotentials[47-48], and the PBE[49] Generalized Gradient Approximation exchange-correlation functional was employed with the plane-wave cutoff energy of 40 Ry for the Kohn-Sham orbitals. For the time-dependent Kohn-Sham equation, the maximum localized Wannier functions (TD-MLWF) gauge was used and time-dependent electric field was applied using the length gauge as discussed in Ref. [24]. The enforced time-reversal symmetry (ETRS) integrator[50] was used with the integration step size of 0.1 a.u. The simulations were performed with the applied electric as the driving field with field period range of 50~400 a.u. and amplitude range of 2~7 × $10^{-3}$ a.u. at uniform intervals of 25 a.u. and 0.5 × $10^{-3}$ a.u., respectively.

## Acknowledgement
This work was supported by the National Science Foundation under Award Nos. CHE-1954894 and OAC-17402204. DCY was supported by an appointment to the Intelligence Community Postdoctoral Research Fellowship Program at Massachusetts Institute of Technology, administered by Oak Ridge Institute for Science and Education through an interagency agreement between the U.S. Department of Energy and the Office of the Director of National Intelligence.

## Supporting Information

The Supporting Information is available free of charge at underline{here.} Determinant of overlap matrix, integrated current, averaged MLWF spreads, time-dependent change of the MLWF spreads, the movement of Wannier function centers, resonance structure schematics of dynamical transition orbital, and molecular geometry. Videos of the dominant dynamical transition orbital changes for $W$=1, 2, and 3.